\documentclass[journal]{IEEEtran}
\ifCLASSINFOpdf
\else
\fi
\usepackage{booktabs}
\usepackage{graphicx}
\usepackage{threeparttable}
\begin{document}

%
\title{Emotion Recognition From Gait Analyses: Current Research and Future Directions}
%
%
%

\author{Shihao~Xu, Jing~Fang, Xiping Hu, Edith~Ngai, Wei Wang, Yi Guo, and Victor C. M. Leung,~\IEEEmembership{Fellow,~IEEE}
\thanks{
\textit{(Shihao Xu and Jing Fang are co-first authors.) (Corresponding authors: Xiping Hu; Wei Wang.)}}
\thanks{S. Xu and X. Hu are with the School of Information Science and Engineering, Lanzhou University, Gansu 730000, China (e-mail: xushh16@lzu.edu.cn; huxp@lzu.edu.cn).}
\thanks{J. Fang is with Shenzhen Institutes of Advanced Technology, Chinese Academy of Sciences, Shenzhen 518055, China (e-mail: jing.fang@siat.ac.cn)}
\thanks{W. Wang is with the School of Intelligent Systems Engineering, Sun Yat-sen University, Shenzhen, Guangdong 518107, (e-mail:  wangw328@mail.sysu.edu.cn).}
\thanks{E. Ngai is with the Department of Electrical and Electronic Engineering, The University of Hong Kong, Hong Kong SAR (e-mail: chngai@eee.hku.hk).}
\thanks{Y. Guo is with the Second Clinical Medical College, Jinan University, Shenzhen 518055, China (e-mail: xuanyi\_guo@163.com). }

\thanks{V. C. M. Leung is with the College of Computer Science and Software Engineering, Shenzhen University, Shenzhen 518055, China, and also with the Department of Electrical and Computer Engineering, the University of British Columbia, Vancouver, B.C. V6T 1Z4, Canada (e-mail: vleung@ieee.org).}

}

%
%

\markboth{}%
{Shell \MakeLowercase{\textit{et al.}}: Bare Demo of IEEEtran.cls for IEEE Journals}
%



\maketitle

\begin{abstract}
Human gait refers to a daily motion that represents not only mobility, but it can also be used to identify the walker by either human observers or computers. Recent studies reveal that gait even conveys information about the walker's emotion. Individuals in different emotion states may show different gait patterns. The mapping between various emotions and gait patterns provides a new source for automated emotion recognition. Compared to traditional emotion detection biometrics, such as facial expression, speech, and physiological parameters, gait is remotely observable, more difficult to imitate, and requires less cooperation from the subject. These advantages make gait a promising source for emotion detection. This article reviews current research on gait-based emotion detection, particularly on how gait parameters can be affected by different emotion states and how the emotion states can be recognized through distinct gait patterns. We focus on the detailed methods and techniques applied in the whole process of emotion recognition: data collection, preprocessing, and classification. At last, we discuss possible future developments of efficient and effective gait-based emotion recognition using the state of the art techniques in intelligent computation and big data.
\end{abstract}


%
\IEEEpeerreviewmaketitle

\section{Introduction}

%
\IEEEPARstart{H}{uman} gait is a manner of walking of individuals. It describes a common but important daily motion through which observers can learn much useful information about the walker. In clinical terms, apart from the detection of movement abnormalities, observation of gait patterns also provides diagnostic clues for multiple neurological disorders such as cerebral palsy, Parkinson's disease, and Rett syndrome \cite{Gage1993,Jellinger1988,Jankovic2008, you2021alzheimer} in an early stage. Clinical gait analysis therefore plays a more and more important role in medical care which may prevent patients from permanent damage. It becomes a well developed tool for quantitative assessment of gait disturbance, which can be applied to functional diagnosis, treatment planning and monitoring of disease progression \cite{Baker2016, giorgi2021using}. In addition, gait provides useful social knowledge to the observers. Research has revealed that human observers are able to recognize themselves and other people that they are familiar with even from the point-light depicted or impoverished gait patterns \cite{Cutting1977,Loula2005}, indicating that gait is unique. The identity information is embedded in the gait signature, which has been considered as a unique biometric identification tool. With the advacements of computer vision and big data analysis, gait recognition has been widely employed for various security applications \cite{Kale2004,Boulgouris2005}.


Furthermore, it was suggested that emotion expression is embedded in the body languages including gait and postural features \cite{Castellano2007,Montepare1987,Coulson2004,Wallbott1998,cicirelli2021human}. Indeed, people in different emotional states show distinct gait kinematics \cite{Gross2012, cai2020feature}. For instance, studies found that depressed individuals show different gait patterns, including slower gaits, smaller stride size, shorter double limb support, and cycle duration, in contrast to the control group \cite{michalak2009embodiment, hu2018emotion}. According to the previous studies, human observers are able to identify emotions based on the gait \cite{Montepare1987}.
These findings indicate that gait can be considered as a potential informative source for emotion perception and recognition. 
\par
Human emotion is heavily involved in cognitive process and social interaction. In recent years, automated emotion recognition becomes a growing field along with the development of computing techniques. It supports numerous applications in inter-personal and human-computer interaction such as customer services, interactive gaming and e-learning, etc.

However, current research on emotion recognition mainly focused on facial expression \cite{Sariyanidi2015,Fasel2003}, speech (including linguistic and acoustic features)\cite{ElAyadi2011} and physiological signals (e.g. electroencephalography, electromyography, heart rate, blood volume pulse, etc.) \cite{Kim2004, moretta2022early, vodovar2022mechanisms, kraus2022prediction}, while relatively few studies investigate the association between emotion and full-body movements. Gait analysis shows apparent advantages over the prevailing modalities, which makes it a promising tool for emotion recognition. We summarize these advantages as follows.

\begin{itemize}
\item Human bodies are relatively large and have multiple degrees of freedom. Unlike facial expression data which must be collected by cameras in very short distance. For monitoring gait patterns, the subjects are allowed to be relatively far from the cameras. The practical distance can be up to several meters while most other biometrics are no longer observable or provide very low resolution \cite{Gunes2009, Walk1988}.
\item Walking usually requires less consciousness and intention from the walker. Thereby it is not vulnerable to active manipulation and imitation. Ekman \cite{Ekman1967} has pointed out that the head expresses the nature of emotions while the body shows information about the emotional intensity.
\item The data collection process of gait patterns requires less cooperation from the subject, so that his or her behaviour is less interfered and closer to the normal state in real life. 
\end{itemize}
\quad With a goal of exploring the opportunities and facilitating the development of this emerging field, we systematically review the literatures related to emotion detection based on gait. A recent survey \cite{avanzino2018relationships} discussed how emotion affects gait for patients with Parkinson disease. Different from that survey, our study focuses on a generic target group with physically and mentally healthy individuals in this work. Another survey by Stephens-Fripp et al. \cite{stephens2017automatic} focused on emotion detection based on  gait and posture, but it discussed gait and posture together and emphasized more on the posture part whereas our current paper is all around gait and shares the details of how gait can be affected by emotions and the process of gait-based emotion detection. Moreover, we also introduce emotion models in current emotion theory and dynamics of gait. It is not only informative, but also essential for computer scientists to integrate theories into their design of the automated emotion recognition system since this field is highly interdisciplinary.

\par This survey is organized as follows: Section \uppercase\expandafter{\romannumeral2} and \uppercase\expandafter{\romannumeral3} give some background on different models of emotion, gait initiation and gait cycle.  Section \uppercase\expandafter{\romannumeral4} discusses the impact of emotion on the gait initiation and the gait cycle. Section \uppercase\expandafter{\romannumeral5} presents the details of gait based emotion recognition including data collection, preprocessing,  and classification. In Section \uppercase\expandafter{\romannumeral6}, we propose some future directions on gait based emotion recognition with machine learning and data analysis. Finally, Section \uppercase\expandafter{\romannumeral7} gives the conclusion on this topic. 
\par

\section{Models of Emotion}
Human emotion is a diverse complex. It can be reflected by a clear happy representation like laughter or smile to extreme sadness with tears. Emotion is hard to be defined comprehensively since it is identified in context-dependent situations. \par
\begin{figure*}[h]

  \centering
  \includegraphics[scale=0.55]{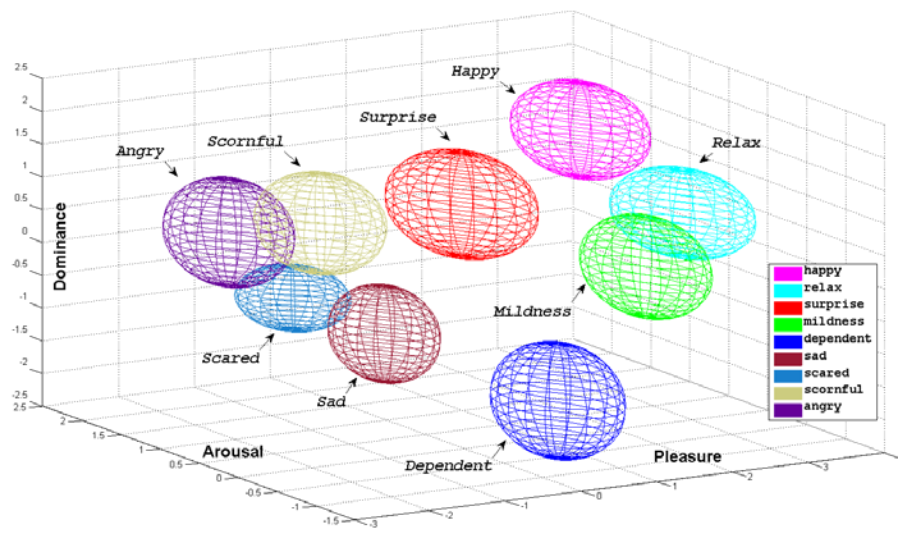}\\
  \caption{The pleasure-arousal-dominance space for emotions\cite{zhang2010facial}\cite{li2005reliability}. The center of each ellipsoid is the mean and the radius is the standard deviation of each emotional state.}\label{emotion_space}
\end{figure*}

There are three common models to describe and measure emotional states in scientific analysis: 1) distinct categories; 2) pleasure-arousal-dominance model (PAD model); and 3) appraisal  approach. The model based on distinct categories  is simple and intuitive, which consists of  six discrete emotions including happiness, sadness, fear, anger, surprise, and disgust. They are often linked to human facial expression \cite{ekman1993facial}. 
The PAD model is a continuous space divided by three orthogonal axes shown in  Fig. \ref{emotion_space}. The pleasure dimension identifies the positivity-negativity of emotional states, the arousal dimension evaluates the degree of physical activity and mental alertness of emotional state, and the dominance dimension indicates to what extend the emotional state is under control or lack of control\cite{zhang2010facial}. Each type of emotion has its place in the three dimensional space including the former mentioned distinct emotions. For instance, sadness is defined as negative pleasure, positive arousal, and negative dominance according to the mapping rules between distinct emotions and the PAD model \cite{mikels2005emotional,morris1995observations}. This model has been used to study nonverbal communication like body language in psychology\cite{mehrabian2017nonverbal}. The third model is appraisal model which describes how emotions develop, influence, and are influenced by interaction with the circumstances\cite{rolls2005emotion,scherer1999appraisal}. Ortony et al. \cite{ortony1990cognitive} applied this appraisal theory to their models and found out that the parameters in the environment such as the events or objects may affect the emotion strength. Because of the complexity of individuals'  evaluations (appraisals or estimates) on events that cause specific reactions in themselves, the appraisal method is less often applied for recognition of emotional states than the two former models. 

\section{Gait}
Human gait refers to a periodical forward task that requires precise cooperation of the neural and musculoskeletal systems to achieve dynamic balance. This requirement is so crucial since the human kinematic system controls the advancing with the support from the two feet. In particular, when only one foot provides standing, the body is in a state of imbalance and it needs a complicated mechanism to make accurate and safe foot trajectory. In the next two subsections, we will talk about the gait initiation and the gait cycle since both of them would be affected by emotion. \par 
\subsection{Gait Initiation}
As an individual starts to move from the static state, he or she is doing gait initiation. In this initiation, anticipatory postural adjustments (APA) plays an important role in breaking the erect posture and transferring the body center of gravity to the stepping foot~\cite{bouisset1987biomechanical}. 
It is a movement involving the swing of a leg forward and leading to imbalance. 
This imbalance is a result of a swift lateral weight shift and is for generating enough advancing power.
\par
\subsection{Gait Cycle}
\begin{figure*}[h]

  \centering
  \includegraphics[scale=0.55]{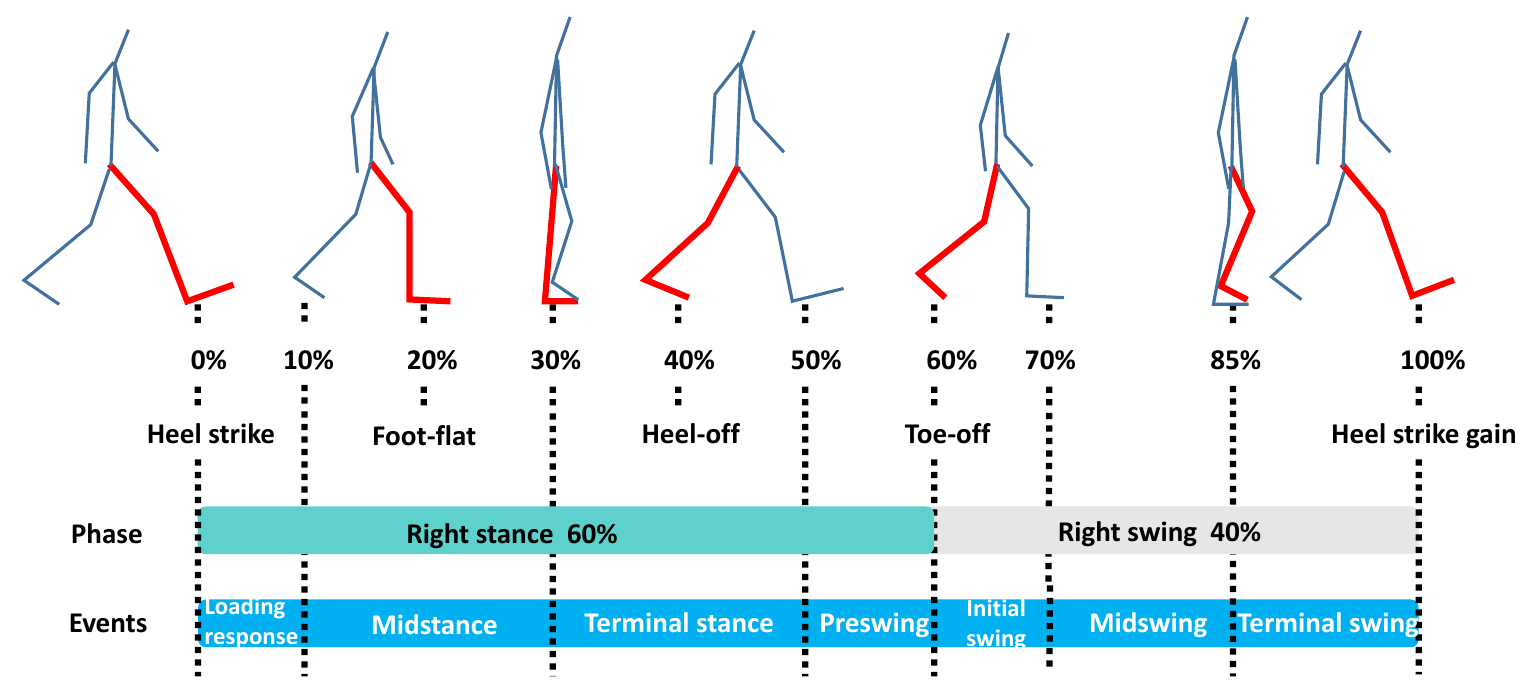}\\
  \caption{Diagram of gait cycle. The right foot (red) shows the gait cycle which can be parted into stance and swing in terms of phase. The stance can be further separated into initial contact, loading response, midstance and preswing. Swing contains initial swing, midswing and terminal swing.}\label{gait}
\end{figure*}

One gait cycle is measured from one heel-strike to another heel-strike with the same heel \cite{vaughan1999dynamics}, which can be described by two cyclic features, phases and events. Both features can be divided into percentage by several key action points (see Fig. \ref{gait}). Two main phases are shown in the gait cycle: In the stance phase, touching the ground  with one foot is called single limb stance.  When both feet are on ground, it is called double support. Following the stance phase, the target foot is going to swing and followed by midswing and terminal swing. Gait can be divided into eight events or subphases. Fig. \ref{gait} shows an example of gait cycle, which starts with initial contact of the right foot in the stance phase. In the initial contact, the crotch extends to a large angle and the knee flexes. Next is the loading response. It is a progress that transfers the body weight from the left leg to the right. Midstance shows the right leg is on the ground whereas the left leg is in motion and the weight is being transferred between them. When the right heel starts to lift and before the left foot lands on the ground, the individual is in terminal stance. Once the left toe touches the ground following the terminal stance, the weight is again uphold by both limbs and the preswing phase begins. Passing through the right toe-off (initial swing) event, the body weight is transferred from the right to the left this time (the midswing phase). As soon as the right heel strikes again, the whole gait cycle is completed and  the walker prepares for next cycle.\par
\begin{figure*}[h]

  \centering
  \includegraphics[scale=0.55]{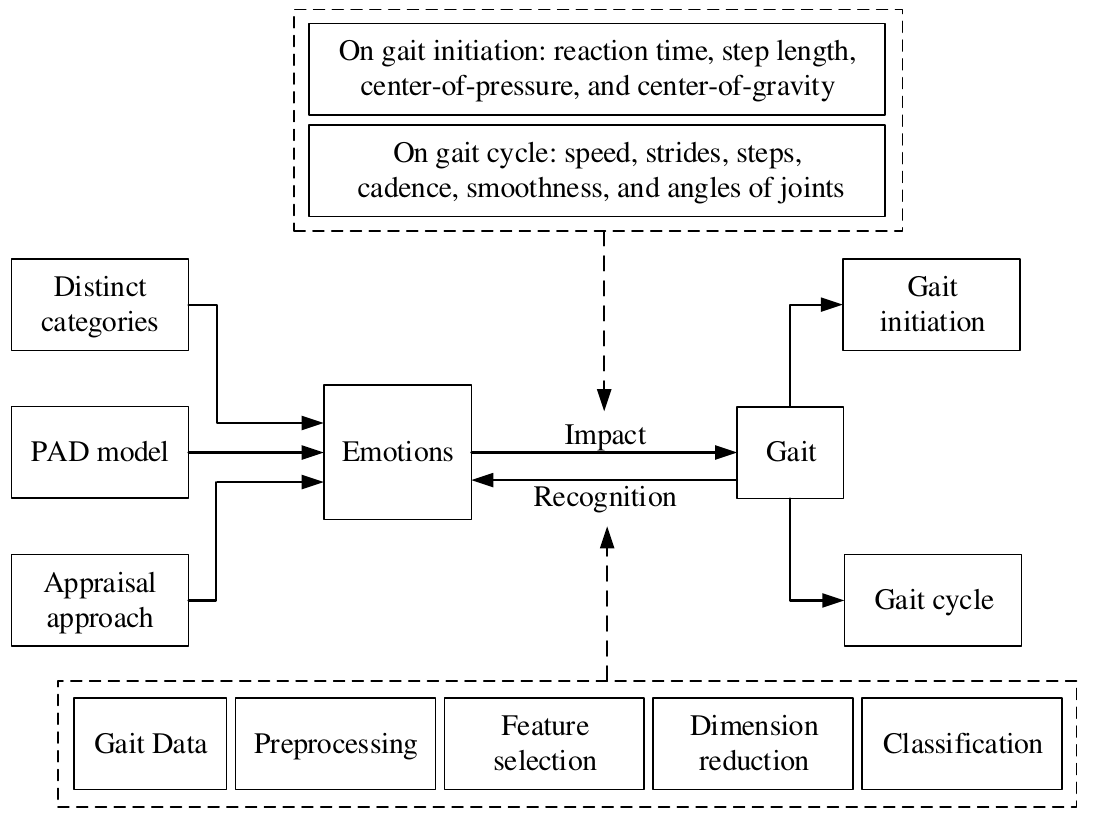}\\
  \caption{Diagram of general relationship between gait and emotion. Three models are used to describe emotion. The gait initiation and cycle can be influenced by emotion. In the opposite, emotion could be recognized through the gait pattern. PAD: Pleasure-arousal-dominance.}\label{general}
\end{figure*}

 
Fig. \ref{general} summarizes the emotion models and the gait cycle that we discussed above. It shows the general relationship between emotion and gait, and gives an overview on the content of this paper. It includes the aforementioned content in Section \uppercase\expandafter{\romannumeral2} and Section \uppercase\expandafter{\romannumeral3}, 
namely the models for describing emotion and the components of 
gait, respectively. It also contains a brief representation of Section \uppercase\expandafter{\romannumeral4} and Section \uppercase\expandafter{\romannumeral5} with the upper part  listing the gait parameters that would be influenced by emotions and the bottom part  showing the process of gait based emotion recognition. 

\section{Emotional Impact on Gait}
\subsection{On Gait Initiation}
Appetitive approach and defensive avoidance are two basic motivational mechanisms for fundamentally organizing emotion in the light of the biphasic theory of emotion \cite{lang2013attention}. Based on this theory,  studies \cite{gelat2011gait,stins2011organization,gelat2015reaction,stins2015biomechanical,naugle2012emotional} have been conducted to explore the emotion effects on gait initiation. Theses studies were usually conducted by giving congruent (CO) tasks and incongruent  (IC) tasks to the walkers. In the CO tasks, the participants were asked to show approaching when sensing the pleasant stimuli or they should express avoidance responding to the unpleasant stimuli. IC tasks were the opposite to the CO tasks as the participants were asked to make approach movements for the unpleasant stimuli or they needed to performed avoidance at the moment they perceived the pleasant stimuli  \cite{chen1999consequences}. Apparently, the IC tasks were related to  the human emotional conflict.  To elicit participants' emotions, the pleasant or unpleasant images from the International Affective Picture System (IAPS) were used as visual stimulus and force plates were used to record the ground reaction forces for the movements. 
\par
We summarize the experimental setups and results from the research studies related to emotion impacts on gait initiation in Table \ref{IC_CO}. In \cite{gelat2011gait}, the specific paradigm was ``go'' or ``nogo'' for volunteers. The ``nogo'' response  (i.e., volunteers were not supposed to move) was related to the stimulus of neutral images showing only an object whereas the ``go'' response (i.e., volunteers should approach or avoid) was for pleasant or unpleasant pictures corresponding to CO or IC tasks. Volunteers were asked to perform responding action as soon as possible after the onset of the images. In \cite{stins2011organization}, participants were asked to make either an anterior step (approach) or a posterior step  (withdrawal) as soon as the image was presented and to remain static till it disappeared. The experiment in \cite{gelat2015reaction} studied the influence of a change in the delay between picture onset and the appearance of ``go'' action for checking people`s reaction time. The changes in the delay had two conditions, namely the short condition  (the word ``go'' showed 500ms after image onset) and the long condition  (the word ``go'' showed 3000ms after image onset). Research\cite{stins2015biomechanical} aimed at gait initiation as soon as the  participants saw the image  (i.e., onset) or as soon as the image disappeared  (offset). In clinical studies, the experiment in \cite{naugle2012emotional} focused on gait initiation in patients with Parkinson's disease 
and the patients were asked to make an initial step with their preferred legs after the image offset and to keep walking at their self-selected pace. 
  \par
Studies mentioned above revealed that an individual's emotion can affect gait initiation. When the participants encounter emotional conflicts, they seemed to pose a defensive response for the IC tasks leading to longer reaction time  (RT) and shorter step length compared to CO tasks shown in Table \ref{IC_CO}. 
This is inspiring because these features learned from gait initiation show significant differences between people with positive and negative emotions. 
The analysis of gait initiation may become a potential method to recognize human's emotion in the future. 

\newcommand{\tabincell}[2]{\begin{tabular}{@{}#1@{}}#2\end{tabular}}
\begin{table*}[h]
\caption{Researches on Impacts of Emotion on Gait Initiation. }
\centering

\begin{tabular}{@{}lllll@{}}
\toprule
Ref& Number of participants& CO tasks & IC tasks& Results \\
\midrule
\cite{gelat2011gait} & \tabincell{ll}{15 (age 20-32,\\9 females) }& \tabincell{ll}{Initiate gait after \\ pleasant image onset} & \tabincell{ll}{Initiate gait after \\ unpleasant image onset} & \tabincell{ll}{Longer RT in IC than CO trials \\and the amplitude of early\\ postural modifications \\was reduced in IC trials. }  \\
\midrule
\cite{stins2011organization}  & \tabincell{ll}{30 (mean age 22.3 years,\\16 females)} & \tabincell{ll}{Approach after \\  pleasant image onset \\or withdraw after \\ unpleasant image onset} & \tabincell{ll}{Approach after \\ unpleasant image onset \\or withdraw after \\pleasant image onset} & \tabincell{ll}{ Unpleasant images caused\\ an initial ``freezing'' response with \\analyses of the preparation,\\initiation,and execution of steps.} \\
\midrule
\cite{gelat2015reaction} & \tabincell{ll}{19 (age 18-26 years, \\11 females) }& \tabincell{ll}{Initiate gait once \\the word ``go''\\ appeared 500 ms \\or 3000 ms after\\ pleasant image onset}& \tabincell{ll}{Initiate gait once \\the word ``go''\\ appeared 500 ms \\or 3000 ms after\\ unpleasant image onset} &  \tabincell{ll}{Motor responses were faster for\\ pleasant pictures than unpleasant ones\\ in the short delay of 500 ms.}\\
\midrule
\cite{stins2015biomechanical} & \tabincell{ll}{27 (mean age 28.7 years, \\16 females)} &\tabincell{ll}{Initiate gait after\\ pleasant image onset or\\ unpleasant image offset}&\tabincell{ll}{Initiate gait after\\ unpleasant image onset \\or pleasant image offset}&  \tabincell{ll}{Gait was initiated faster \\with pleasant images at onset and \\faster with unpleasant images\\ at offset with analyses of COP and COG.}\\
\midrule
\cite{naugle2012emotional}    &  \tabincell{ll}{26 patients \\(age 55-80 years,3 females) \\ 25 normals \\(age 55-80 years,3 females)}& \tabincell{ll}{Initiate gait and walk \\after approach-oriented\\ emotional picture onset}&\tabincell{ll}{Initiate gait and walk \\after withdrawal-oriented\\ emotional picture onset}& \tabincell{ll}{For PD patients and healthy older adults, \\threatening pictures speeded the GI\\ and approach-oriented emotional pictures, \\compared to withdrawal-oriented pictures, \\facilitated the anticipatory postural \\adjustments of gait initiation with analyses\\ of RT and COP.}\\
\bottomrule

\end{tabular}
      \begin{tablenotes}
        \footnotesize
        \item[1] CO: Congruent. IC: Incongurent. RT: Reaction time. COP: Center-of-pressure. COG: Center-of-gravity. GI: Gait initiation. PD: Parkinson's disease.
      \end{tablenotes}

 \label{IC_CO}
\end{table*}

\subsection{On Gait Cycle}
In this subsection, a few studies would be shared for showing the performances and characteristics of emotive gaits (see Table \ref{on_cycle}). In Montepare's investigation\cite{Montepare1987The}, ten female observers viewed the gaits of five walkers with four various emotions (i.e., sadness, anger, happiness, and pride) in order to determine the walkers' emotions and report specific gait features observed. Note that the walkers' heads were not recorded in the gaits to prevent the facial confusion of emotion perception. The investigation showed that gait patterns with different emotions could be identified better than chance level with mean accuracy of 56\%, 90\%, 74\%, 94\% for pride, anger, happiness, and sadness respectively. As for the gait features which differentiated emotions, the angry gaits were relatively more heavyfooted than the other gaits, while the sad gaits had less arm swing compared with the other gaits. It also turned out that proud and angry gaits had longer stride lengths than happy or sad gaits. Finally, happy gaits performed faster in pace than the other gaits. 
\par
Similarly, thirty observers used Effort-Shape method \cite{Laban1971The} to rate the qualitative gait movements of sixteen walkers expressing five emotions (i.e., joy, contentment, anger, sadness, and neutral) in Gross's work \cite{Gross2012Effort}. The Effort-Shape analysis involved four factors evaluating the effort in the walker's movements (i.e., space, time, energy, flow) and two factors described the shape of the body (i.e., torso shape and limb shape) which are shown in  Table \ref{effort shape}. Each factor was rated from 1 to 5. For the instance of flow, the left anchor was ``free, relaxed, uncontrolled'' and the right anchor was ``Bound, tense, controlled''. The three intermediate points in the scale acted a transition between the left and right anchor qualities. Results revealed that the sad gait was featured as having a contracted torso shape, contracted limb shape, indirect space, light energy, and slow time. The angry gait was regarded as having expanded limb shape, direct space, forceful energy, fast time, and tense flow. The joy gait had common characteristics with the angry gait in limb shape, energy, and time, but there was a more expanded torso shape and more relaxed flow for the joy gait than the anger gait. The content gait might look like the joy gait but the former one had more contracted limb shape, less direct space, lighter energy and slower time than the latter one. When it came to the gait pattern in neutral emotion state, it was similar to the sad gait, however, it had a more expanded torso and limb shapes, more direct space, more forceful energy and faster time than the sad gait.

\begin{table*}[h]
\caption{Qualities associated with Effort-Shape factors\cite{Gross2012Effort}.}
\centering
\begin{tabular}{@{}lll@{}}
\toprule
Effort-Shape factor & Left-anchor qualities$^a$          & Right-anchor qualities$^b$        \\ \midrule
Torso Shape         & Contracted, bowed, shrinking    & Expanded, stretched, growing   \\
Limb Shape          & Moves close to body, contracted & Moves away from body, expanded \\
Space               & Indirect, wandering, diffuse    & Direct, focused, channeled     \\
Energy              & Light, delicate, buoyant        & Strong, forceful, powerful     \\
Time                & Sustained, leisurely, slow      & Sudden, hurried, fast          \\
Flow                & Free, relaxed, uncontrolled     & Bound, tense, controlled       \\ \bottomrule
$^a$ Score = 1.\\
$^b$ Score = 5.
\end{tabular}
 \label{effort shape}
\end{table*}
\par
Another report from human observers for perceiving walkers' emotions showed that there were significant differences in gait patterns among people with various emotions such as happiness with a bouncing gait, sadness with a slow slouching pace, anger with a fast stomping gait and fear with fast and short gait \cite{halovic2018not}.\par
Through data analytics, 
the features in gaits carrying various emotions could be studied through kinematic analysis based on the gait data. In  \cite{roether2009critical}, two types of features, 1) posture features and 2) movement features, had been explored to determine which features were the most important in various emotional expression through analyzing the gait trajectory data. For both types of features mentioned above, flexion angles of eleven major joints (head, spine, pelvis and left and right shoulder, elbow, hip and knee joints) have been averaged over the gait cycle. In terms of the posture features, the evident results were the reduced head angle for sad walking, and increased elbow angles for fear and anger while walking. As for movement features, the happy and angry gaits were linked to increased joint amplitudes whereas sadness and fear showed a reduction in joint-angle amplitudes. In addition, the study compared the emotional gaits with neutral gaits whose speeds were matched to the former ones (with overall velocity difference $<$ 15\%) and it figured out that the dynamics of the emotion-specific  features cannot be explained by changes in gait velocity. 
In Barliya's study \cite{barliya2013expression}, the gait speed was shown to be a crucial parameter that would be affected by various emotions. The amplitude of thigh elevation angles differed from those in the neutral gait for all affections except sadness. Anger was showing more frequently oriented intersegmental plane than others. \par
Through the kinematic analysis of emotional (i.e., happy, angry, fearful, and sad ) point-light walkers, Halovic and Kroos \cite{halovic2018not} found out that both happy and angry gaits showed long strides with increased arm movement but angry strides had a faster cadence. Walkers feeling fear were with fast and short strides. Sad walkers had slow short strides gaits representing the slowest walking pace. The fearful and sad gaits both had less arm movement but the former one mainly moved their lower arms whilst the latter one had the entire arms movement. \par
Studies \cite{kang2015emotional, kang2016effect11} applying eight-camera optoelectronic motion capture system focused on smoothness of linear and angular movements of body and limbs  to explore the emotive impact on gaits. In the vertical direction, the smoothness of movements increased with angry and joyful emotions in the whole body center-of-mass, head, thorax and pelvis compared to sadness. In the anterior-posterior direction, neutral, angry, and joyful emotions only had increased  movement smoothness for the head compared to sadness. In angular movements, anger's movement smoothness in the hip and ankle increased compared to that of sadness.  Smoothness in the shoulder increased for anger and joy emotions compared to sadness. \par
Research in \cite{michalak2009embodiment,lemke2000spatiotemporal} further analyzed the gaits of patients with depression compared wtih the control group. It found that depressed patients showed lower velocity, reduced limb swing and less vertical head movements.  
\begin{table*}[]
\caption{Researches on Impacts of Emotion on Gait Cycle.}
\centering
\scalebox{1}{
\setlength{\tabcolsep}{0.9mm}{
\begin{tabular}{@{}lllll@{}}
\toprule
Ref & Emotion & Number of walkers & Methods & Performances  \\
\midrule

\cite{Montepare1987The}& \tabincell{ll}{Happiness,sadness,\\anger,and pride}  & \tabincell{ll}{10 females}& \tabincell{ll}{5 female observers} & \tabincell{ll}{Mean accuracy: 56\%,90\%, 74\%, 94\% \\for pride, anger, happiness, and sadness respectively\\
Anger: heavyfooted.\\ Sadness: less arm swing. \\
Proud and anger: longer stride lengths. \\Happiness: faster.} \\
\midrule

\cite{Gross2012Effort}& \tabincell{ll}{Joy,contentment,\\sadness,anger,\\and neutrality} & \tabincell{ll}{11 females and 5 males}& \tabincell{ll}{30 (15 females) observers \\with Effort-Shape method} & \tabincell{ll}{
Sadness: contracted torso shape, \\contracted limb shape, indirect space,\\
light energy, and slow time. \\
Anger: expanded limb shape, \\direct space, forceful energy,\\
fast time, and tense flow. \\
Joy: more expanded torso shape \\
and more relaxed than anger.\\
Content: more contracted limb shape, \\
less direct space,lighter energy \\
and slower time than the joy. \\
Neutrality: more expanded torso \\
and limb shapes, more direct space,\\ 
more forceful energy\,\ and faster time than sadness.}  \\
\midrule

\cite{halovic2018not}& \tabincell{ll}{Happiness,sadness,\\anger,fear,and neutrality}& \tabincell{ll}{36 actors (17 females)} & \tabincell{ll}{34 (19 females) observers\\and the kinematic analysis} & \tabincell{ll}{
Happiness and anger: long strides \\with increased arm movement\\ but angry strides had a faster cadence.\\ Fear: fast and short strides. \\Sad: slowest short strides gaits.\\ Fear and sadness: less arm movement\\ but fear  mainly moved their lower arms \\whilst sadness had the entire arms movement}\\

\midrule

\cite{roether2009critical}& \tabincell{ll}{Happiness,sadness,\\anger,fear,and neutrality}& \tabincell{ll}{11 males and 12 females} & \tabincell{ll}{Average flexion angles,\\ nonlinear mixture model,\\and sparse regression} & \tabincell{ll}{
Posture  features were  the  \\reduced  head  angle  for  sad  walking,  \\and increased elbow angles for fear and anger.\\
Movement features were increased joint amplitudes \\for happiness and anger, and \\ a reduction in joint-angle amplitudes\\ for sadness and fear.}  \\
\midrule
\cite{barliya2013expression}& \tabincell{ll}{Happiness, sadness,\\anger,fear,and neutrality }& \tabincell{ll}{13 university students \\and 8 professional actors} & \tabincell{ll}{The intersegmental \\law of coordination} & \tabincell{ll}{Speed was affected by emotions.\\The amplitude of thigh elevation angles \\differed from those in neutral gait\\ for all emotions except sadness.\\ Anger was showing more\\ frequently oriented intersegmental\\ plane than others.}  \\

\midrule
\cite{kang2016effect11}& \tabincell{ll}{Joy, sadness,\\anger,and neutrality }& \tabincell{ll}{7 males and 11 females} & \tabincell{ll}{Measuring spatiotemporal\\ gait parameters \\and smoothness of \\linear movements} & \tabincell{ll}{ In the VT direction, angry and joyful
movement\\ smoothness increased compared to sadness. \\
In the AP direction, neutral, angry, \\and joyful gaits had increased movement \\smoothness for the head compared to sadness. \\In angular movements, anger's smoothness in\\
the hip and ankle increased compared to sadness.\\
Smoothness in the shoulder \\increased for anger and joy
compared to sadness.}  \\

\midrule
\cite{michalak2009embodiment}& \tabincell{ll}{Depression \\and non-depression}& \tabincell{ll}{14 inpatients \\with major depression \\ and 14 healthy people} & \tabincell{ll}{Analysing\\ forward/backward\\ movements,
VT movements,\\ and lateral movements \\of all body segments } & \tabincell{ll}{Depression: reduced walking \\
speed, arm swing, and vertical head movements}  \\

\midrule
\cite{lemke2000spatiotemporal}& \tabincell{ll}{Depression \\and non-depression}& \tabincell{ll}{16 inpatients \\with major depression \\ and 16 healthy people} & \tabincell{ll}{Measuring spatiotemporal\\ gait parameters } & \tabincell{ll}{Depression: lower gait velocity,\\ reduced stride length, \\double limb support and cycle duration.}  \\

\bottomrule

\end{tabular}}}
 \begin{tablenotes}
        \footnotesize
        \item[1]VT: Vertical. AP: Anterior-posterior.
      \end{tablenotes}

 \label{on_cycle}
\end{table*}

 \begin{figure}[h]

  \centering
  \includegraphics[scale=0.4]{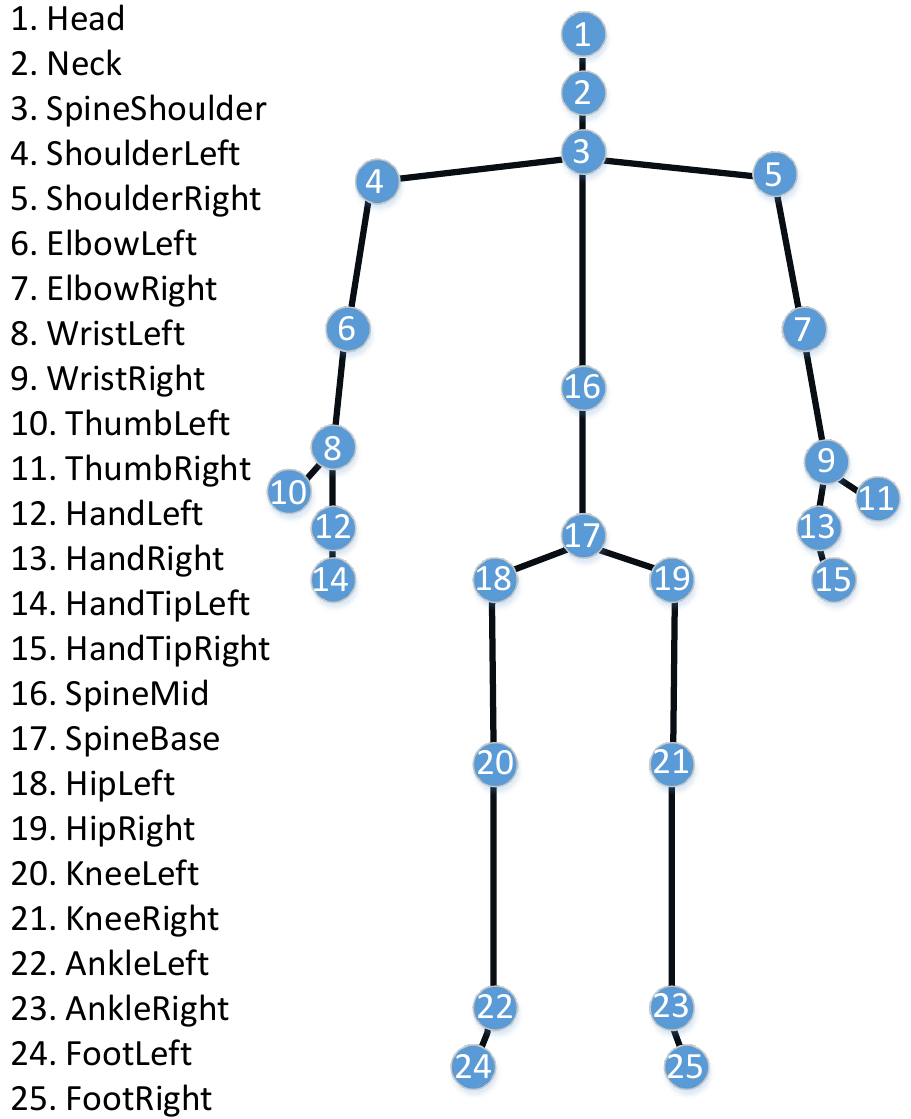}\\
  \caption{\quad 25 joints of human generated by Kinect  }\label{skeleton}
\end{figure}

\section{Gait Analysis for Emotion Recognition}
There is a long way to go towards the ultimate goal, namely, automated emotion recognition through gait patterns. However, there are important observations based on previous studies~\cite{wang2019towards, sheng2021multi, kanjo2019deep}. Automated process may be possible by aggregating data from previous observations supported by big data analysis and machine learning. In this subsection, we discuss some details of current research on gait based emotion recognition.\par

\subsection{Gait Data Collection}
Gait can be captured in digital format (i.e. by cameras, force plates) for data analysis. As the technology is advancing, digital devices have increased sensing quality and become more intelligent. In the field of emotion detection based on gait, the force platform was very useful for recording the anteroposterior velocity and displacement of center of foot pressure\cite{janssen2008recognition}. The infrared light barrier system functioned as well for measuring gait velocity \cite{lemke2000spatiotemporal,janssen2008recognition}. 
More prevailing was the motion analysis systems  (e.g., Vicon and Helen Hayes) that could capture precise coordinates information of markers by taping them over people 
\cite{roether2009critical,barliya2013expression,omlor2007extraction,michalak2009embodiment,Gross2012Effort,destephe2013influences,venture2014recognizing,karg2010recognition}. As another convenient and non-wearable device, Microsoft Kinect, showed up. It was first for interactive games, and then it could be used for motion analysis due to its representation of human skeleton \cite{li2016identifying,li2016emotion}. 
Fig. \ref{skeleton} shows an individual's skeleton consisting of marking 25 joints, which were displayed by Kinect V2 device. The joints were figured out based on the three dimension coordinates derived from depth image. In addition, Kinect can provide other types of video, such as  RGB video and infrared video~\cite{wang2020gait,lu2021new}. In recent years,  intelligent wearable devices got a lot of attention not only in the market, but they could be used in gait pattern analysis for identifying people's emotion since the accelerometer in the devices collect movement data \cite{zhang2016emotion,quiroz2018emotion,9623511}. In the work of Chiu \cite{chiu2018emotion}, the mobile phone was used for capturing the gait of people  and then the video data was sent to a server for pose estimation and emotion recognition.

\subsection{Preprocessing}
Before feeding data to the classifiers or performing correlation analysis, the raw data should be preprocessed to obtain significant and concise features for computation efficiency and performance. Data preprocessing is a crucial step that may include many substeps such as information filtering that helps to remove noise artifacts and burrs and to perform data transformation,  feature selection, and so on. In the following paragraph, we will present some of the preprocessing techniques.
\begin{enumerate}
\item Filtering \par
To smooth the data and get rid of the noise for the marker trajectories of an individual's walking, studies \cite{kang2015emotional,kang2016effect11,Gross2012Effort,destephe2013influences} used low-pass Butterworth filter with a cut-off frequency of 6 Hz. Butterworth filter is considered as a maximally flat magnitude filter \cite{butterworth1930theory} since the frequency response in the passband has no ripples and rolls off towards zero in the stopband \cite{bianchi2007electronic}. 


\quad Sliding window Gaussian filtering is another common way to eliminate some noises or high-frequency components like jitters. Mathematically, a discrete Gaussian filter transfers the input signal by convolution with a Gaussian function in which there is a significant parameter called standard deviation and the standard deviation is key to designing a Gaussian kernel of fixed length. In studies \cite{li2016identifying,li2016emotion}, the Gaussian kernel was $\frac{1}{16}[1,4,6,4,1]$ for filtering  three dimensional  coordinates of joints of walkers .


\item Data Transformation
\par
Processing the data in the time domain may not be the most effective method. In most of the time, data transformation to other domains, like the frequency domain or time-frequency domain, is favorable, which can make the understanding of the data more thorough.
\par
\quad One classical and popular method is the discrete Fourier transform (DFT)  \cite{proakis2001digital}.
Discrete Fourier transform derives frequency information of the data, providing the Fourier coefficients that can feature the data. This transform can be applied in  three dimensional gait data recorded by Microsoft Kinect devices\cite{li2016identifying,li2016emotion,sun2017self}\par
\quad Another method outweighing discrete Fourier transform is the discrete wavelet transform (DWT) that  represents the frequency and location (location in time) information, which has been applied in gait studies \cite{baratin2015wavelet,ismail1999discrete,nyan2006classification}. In DWT, selecting a wavelet that best suited the analysis is crucial and the choice of the best wavelet is generally based on the  the signal characteristics and the applications. For example, for data compression, the wavelet is supposed to represent the largest amount of information  with as few coefficients as possible. 
Many wavelets have been proposed ranging from the Haar, Daubechies, Coiflet, Symmlet, and Mexican Hat to Morlet wavelets and they possess various properties that meet with the needs of the work. For instance, the Daubechies 4 (Db4) is for dealing with signals that have linear approximation over four samples whereas Db6 aims at quadratic approximations over six samples \cite{mallat1992singularity}. 

\item Feature Selection\par
Investigations have been conducted to provide qualitative and quantitative evaluations of multiple features in the human gait. The way to choose the parameters of interest depends on specific application domain. As an example in sports, Electromyography (EMG) signal recorded from muscles was exploited to build a model for determining muscle force-time histories when an individual was walking\cite{white1992predicting}.
\begin{itemize}
\item Spatiotemporal Features\par
The gait cycle contains many useful spatiotemporal quantities such as gait velocity, stride and step lengths, step width, single/double support and swing period, phases, rhythm  (number of steps per time unit), foot placement through the measurement of time and scale.

\item Kinematic Features \par
Gait data of the joints or skeleton collected by marker trajectories, Kinect cameras or video-based pose estimation, are helpful for the measurements of kinematic features, which are not limited to joint angles, angular range of motion, displacement and velocity on basis of various axes\cite{barliya2013expression}. Research in \cite{roether2009critical} explored the crucial kinematic features related to emotion from gait. It indicated that limb flexion is a key feature for expression of anger and fear on gait whereas head inclination corresponded to sadness dominantly. More specific adoption of kinematic features could be found in the next section.  

\end{itemize}

\end{enumerate}

\subsection{Emotion Recognition}
There have been several efforts to classify a walker's emotion through their gait data (see Table \ref{acc_table}). In \cite{janssen2008recognition}, Janssen et al. used the Kistler force platform to record the ground reaction forces of walkers who were sad, angry or happy through recalling specific occasion when they felt the corresponding emotion. 
After a short period of elicitation of various emotions, volunteers were asked to walk about 7m at a self-determined gait velocity on the force platform. The ground forces of gait after normalization by amplitude and time were separated into two parts, training set and test set. The training part was put into a supervised Multilayer Perceptrons (MLP) with 200-111-22 (input-hidden-output) neurons (one output neuron per participant). The test part was for the validation that MLP model achieved 95.3\% accuracy of recognizing each individual. With the help of unsupervised Self-organizing Maps (SOM), the average emotion recognition rate was 80.8\%.\par
Omlor et al. \cite{omlor2007extraction} tried to classify emotions expressed by walkers with attached markers upon joints. They presented an original non-linear source separation method that efficiently dealt with temporal delays of signals. This technique showed superiority to PCA and Independent Component Correlation Algorithm (ICA). The combination of this method and sparse multivariate regression figured out spatio-temporal primitives that were specific for different emotions in gait. The stunning part is that this method approximated movement trajectories very accurately up to 97\% based on three learned spatio-temporal source signals.\par
In \cite{venture2014recognizing,5626404}, feature vectorization (i.e computing the auto-correlation matrix) has been applied in gait trajectories. Four professional actors feeling neutral, joy, anger, sadness, and fear respectively repeated five times walks in a straight line with joint-attached 41 markers of a Vicon motion capture system. In the vector analysis, the angles, position, velocity, and acceleration of joints have been explored for detecting emotions. The study found out that lower torso motion, waist, and head angles could significantly characterize the emotion expression of walkers whereas the leg and arm biased the detection. More importantly with the utilization of the weight on vector, the emotion recognition has been improved to a total average accuracy of 78\% for a given volunteer. 
\par
\begin{figure}[h]
  \centering
  \includegraphics[scale=0.5]{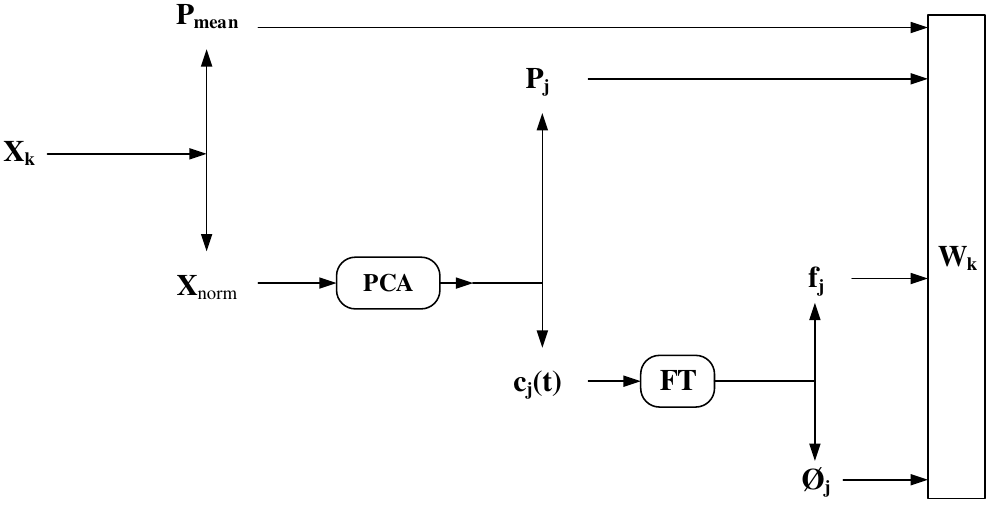}\\
  \caption{Component description of $W_{k}$. It consists of the mean posture $P_{mean}$, four eigenpostures $P_{j}$, four frequencies $f_{j}$, and three phase shifts  $\phi_{j}$. \cite{karg2010recognition}} \label{pca_ft_pca} 
\end{figure}

 
Karg et al. studied thoroughly affect recognition based on gait patterns \cite{karg2010recognition}. Statistical parameters of joint angle trajectories and modeling joint angle trajectories by eigenpostures were two ways to extract features in the research. The former consists of three parts. The first part was calculating the velocity, stride length and cadence  (VSC). The second part was figuring minimum, mean, and maximum of significant joint angles (i.e., neck angle, shoulder angle, thorax angle) including VSC whereas the third part applied the same method as the one in the second part to all joint angles. These latter two parts were processed independently by PCA, kernel PCA (KPCA),  linear discriminant analysis (LDA), and general discriminant analysis (GDA) respectively. The other way based on eigenpostures was to establish the matrix $W$ that contained the mean posture, four eigenpostures, four frequencies and three phase shifts  (later referred to as PCA-FT-PCA) as shown in Fig. \ref{pca_ft_pca}. Given two kinds of features above, for recognition, the next step was to feed them to the classifiers Naive Bayes, Nearest Neighbor  (NN) and SVM respectively. The essence of this research was to further explore the interindividual and person-dependent recognition as well as the recognition based on the PAD  (pleasure, arousal, dominance) model\cite{russell1977evidence}. The best accuracy of affect recognition was achieved, 93\%, based on the estimated identity as shown in Table \ref{acc_table}.\par
\begin{figure}[h]
  \centering
  \includegraphics[scale=0.58]{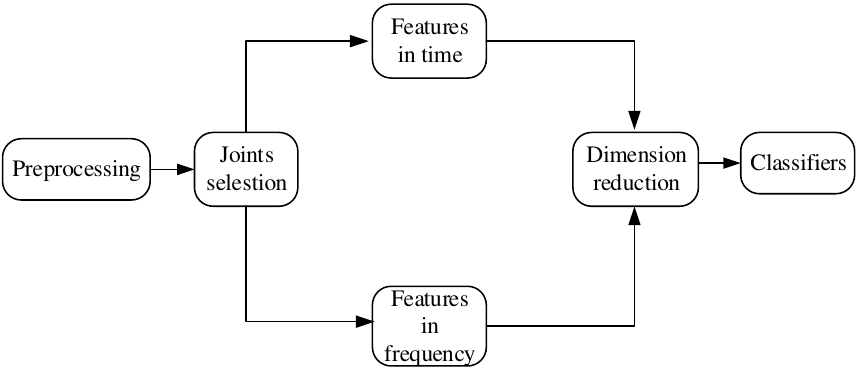}\\
  \caption{Procedure of emotion recognition based on Kinect. Joints selections includes 2 wrists, 2 knees and 2 ankles coordinates. Features in time domain were mean and variance of stride, period, velocity, and height whereas frequency features were the amplitudes and phases of top 20 frequencies. } \label{kinect_process}
\end{figure}

Research in \cite{li2016identifying,li2016emotion} both used Microsoft Kinect v2 sensor, a low-cost and portable sensor to recognize emotional state through people's 1-minute gaits in straight line walking back and forth after they watched a 3-minute video clips that may cause various emotion elicitation. There were 59 actors experiencing happiness, anger, and neutral state respectively. 
Kinect camera captured the gait data in the format of 3D coordinate of 25 joints. After data preprocessing (data segmentation, low-pass filtering, coordinates translation and coordinate difference), the selected joints were then featured in the time and frequency domain. After that they were finally fed into classifiers including the LDA, Naive Bayes, Decision Tree and SVM respectively\cite{li2016identifying} (see Fig. \ref{kinect_process}). The distinct improvements were achieved by PCA with selecting the useful major features that the highest accuracy were 88\% using SVM for happiness between happiness and neutral, 80\% using Naive Bayes for neutral between anger and neutral and 83\% using SVM for anger between happiness and anger whilst Li et al. \cite{li2016emotion} used Naive Bayes, Random Forest, SVM and SMO respectively to the skeletons data and the highest accuracy were 80.51\% , 79.66\%, 55.08\% with Naive Bayes (for anger and neutral) , Naive Bayes  (for happiness and neutral) , Random Forests (for happiness and anger) accordingly. \par

In \cite{zhang2016emotion}, gait pattern was tracked by a customized smart bracelet with built-in accelerometer, which recorded three dimensional acceleration data from different body positions such as right wrist and ankle. In order to remove unexpected waking vibrations and reduce data redundancy, they preprocessed the time series data by using moving average filter and sliding window. Then, with the calculation of skewness, kurtosis and standard deviation of each one of three axes, the temporal domain features were selected. What is more, they  used Power Spectral Density  (PSD) and Fast Fourier Transform  (FFT) to extract temporal and frequency domain features. After this part, the final number of features was 114 (38 features on each axis), and they were fed into 4 different algorithms  (Decision Tree, SVM, Random Forest, and Random Tree) respectively. In the eventual result, SVM achieved the highest accuracy of classification,  88.5 \%, 91.3\%, 88.5\%, respectively for happiness and neutral, anger and neutral, happiness and anger. Results indicated that it was efficient to recognize human emotions (happiness, neutral and angry) with gait data recorded by a wearable device. \par

In the investigation by Juan et al. \cite{quiroz2018emotion}, the accelerometer data from the walker`s smart watch has been used to identify emotion. Fifty participants were divided into groups to perform three tasks: 1)watch movies and walk 2)listen to music and walk 3) listen to music while walking. Movies or music were for eliciting participants happiness or sadness for the later tasks. Either after stimulation or engaging in it, volunteers walked in a 250m S-shaped corridor in a round trip while wearing a smart watch . When it comes to feature selection, the accelerometer data would be divided by sliding windows and features came from each window as a vector. Then Random forest and Logistic Regression handled the classification for happiness and sadness based on the feature vectors, and the majority of accuracies ranged from 60\% to 80\% compared to the baseline 50\%.
\par Chiu et al. aimed at emotion recognition using mobile devices\cite{chiu2018emotion}. Eleven male undergraduate students were recruited to show emotive walks feeling five emotions (i.e., joy, anger, sadness, relaxation, and neutrality) with about ten seconds for each walk when a mobile phone camera was recording from the left side of the subject. After data collection, videos were fed into OpenPose model to extract 18 joints positions in pixel xy-coordinate format for each frame. Then three main features 1) euclidean features 2)angular features and 3) speed were calculated. They were euclidean distances  between joint positions (i.e., two hands, left hand and left hip, two knees, left knee and left hip, two elbows, two feet) normalized by the height of the bounding box surrounding subject in pixels and angles of joint positions (i.e., two arms, left and right elbow flexion, two legs, left and right knee flexion, vertical angle of the front thigh, angle between front thigh and torso, vertical angle of head inclination, angle between the head and torso) as well as  several types of speed (i.e., average total speed, average each step speed, standard deviation  of each step speed, maximum and minimum of each step speed). Then features were used for training six models respectively (i.e., SVM, Multi-layer Perceptron, Decision Tree, Naive Bayes, Random Forest, and Logistic Regression). All the computations were done in a server and the result would be sent back to the client side, the mobile phone. Evaluation results presented that single SVM achieved the highest accuracy of 62.1\% in classification compared with other classifiers, while human observers achieved 72\% accuracy. 
\par
In another research of gait based emotion recognition\cite{ahmed2018score}, seven subjects' five emotive gaits (i.e., happiness, sadness, anger, fear, neutrality) with 20 seconds duration in each walking sequence were recorded in skeleton format by Kinect v2.0 camera. Geometric and kinematic features were calculated using Laban Movement Analysis (LMA) \cite{Laban1971The} and those features acted as LMA components from four aspects Body, Effort, Shape, and Space. Then the binary chromosome based genetic algorithm was adopted to determine a subset of features that gained the maximum accuracy of four classifiers (i.e., SVM, KNN, Decision Tree, LDA) for emotion recognition. Furthermore, the four classifiers were combined with Score-level and Rank-level Fusion algorithms to boost the overall performance. At last the Score-level Fusion method helped  to achieve 80\% emotion recognition accuracy, which outperformed any single one of those four models. 

More recently, Shao et al.~\cite{9585554} claimed that detecting people at risk of depression in a non-contact and effective method is important. For this purpose, they designed a multi-modal gait analysis-based depression detection method that considers both skeleton modality and silhouette modality. In their method, a skeleton feature set including spatiotemporal features and kinematics features are adopted for depression description. Similarly, Lu et al.~\cite{9674794} proposed an integrated gait assessment framework to assess the risk of depression based on the collection and analysis of multimedia data. Specifically, they represented the rigid body based on kinetic energy and potential energy. Then,  the fast Fourier transform is used to analyze these two energies based on the frequency domain. To improve the efficiency of the gait-based method, Rao et al. explored the person re-identification issue via gait features within 3D skeleton sequences using a new self-supervised learning paradigm~\cite{9466418, rao2021self, rao2021sm}. 

\begin{table*}[]
\caption{Studies on emotion recognition based on gait. }
\centering
\scalebox{1}{
\setlength{\tabcolsep}{0.9mm}{
\begin{tabular}{@{}llllll@{}}
\toprule
Ref &Emotions& Features&Method/Classifier & Accuracy/Approximation \\
\midrule
\cite{janssen2008recognition} &  \tabincell{ll}{Happiness,Sadness,\\Anger,Normality}&\tabincell{llll}{Ground reaction force:\\176 gait patterns of 22 actors;\\The force in x,y, and z dimensions \\normalized by amplitude and time} & \tabincell{llll}{Multilayer Perceptron,\\ Self-organizing Maps}& \tabincell{ll}{80.8\% \\for all emotions recognition} \\
\midrule
\cite{omlor2007extraction} & \tabincell{ll}{Happiness,Sadness, \\ Anger,Fear,Neutrality}&  \tabincell{llll}{Marker trajectories: \\195 gait trajectories from 13 lay actors;\\Flexion angles of the hip,knee,elbow,\\shoulder and the clavicle}& \tabincell{ll}{Original non-linear \\source separation,\\Sparse \\Multivariate Regression}& \tabincell{ll}{97\% for describing original data \\with  only 3 source signals}\\ 
\midrule
\cite{venture2014recognizing} & \tabincell{ll}{Joy,Sadness, \\ Anger,Fear,Neutrality}&  \tabincell{llll}{Marker trajectories: \\100 gait trajectories from\\ 4 professional actors;\\Angles,position,velocity,and acceleration \\of 34 joints and generalized coordinates\\ of lower torso}& \tabincell{ll}{Similarity index}& \tabincell{ll}{ 78\% for all emotions recognition \\of an individual}\\ 
\midrule
\cite{karg2010recognition} & \tabincell{ll}{Happiness,Sadness, \\ Anger,Neutrality}&  \tabincell{llll}{Marker trajectories: \\520 gait trajectories of 13 male actors;\\Velocity, stride length
, cadence, \\statistical parameters \\of joint angle trajectories,\\and modeling joint angle trajectories \\by eigenpostures} & \tabincell{ll}{KNN,\\Naive Bayes,\\SVM}& \tabincell{ll}{Highest accuracy for interindividual: \\69\% (SVM for all joint angles + PCA); \\Highest avg accuracy \\for person-dependent: \\95\% (SVM for all joint angles + PCA);\\Highest accuracy based on \\the estimated identity: 92\%\\  (Naive Bayes for all joint angles + PCA);\\Highest accuracy for affective dimensions: \\97\% for Arousal(KNN for all joint angles)
}\\ 
\midrule
\cite{li2016identifying} & \tabincell{ll}{Happiness,Anger \\ Neutrality} & \tabincell{lll}{Kinect skeletons:\\ 59 actors experience\\ each emotion respectively;\\ 6 joints on arms and legs \\in the time and frequency domain}&\tabincell{ll}{LDA,\\Naive Bayes,\\SVM \\with PCA features} & \tabincell{llll}{Highest accuracy:\\ 88\% (SVM for Happiness\\ between Happiness and Neutral),\\80\%  (Naive Bayes for Neutral \\between Anger and Neutral),\\ 83\%  (SVM for Anger \\between Happiness and Anger) }\\
\midrule
\cite{li2016emotion} & \tabincell{ll}{Happiness,Anger\\Neutrality} & \tabincell{llll}{Kinect skeletons:\\59 actors experience\\ each emotion respectively;\\ 42 main frequencies and \\42 corresponding phases of 14 joints\\} &\tabincell{ll}{Naive Bayes,\\Random Forest,\\SVM,\\SMO } & \tabincell{llll}{Highest accuracy:\\ 80.51\%  (Naive Bayes for \\Anger and Neutral),\\79.66\%  (Naive Bayes for\\Happiness and Neutral),\\ 55.08\%  (Random Forest for\\Happiness and Anger)}\\
\midrule
\cite{zhang2016emotion} & \tabincell{ll}{Happiness,Anger,\\Neutrality }&  \tabincell{llll}{Accelerometer data recorded \\by smart bracelet:\\123 actors  experience \\each emotion respectively;\\3D acceleration data recorded \\on right wrist and ankle \\in the time and frequency domain} & \tabincell{ll}{Decision Tree,\\SVM,\\Random Forest,\\Random Tree\\with PCA features}& \tabincell{ll}{Highest accuracy:\\88.5\% (SVM for  Happiness and Neutral),\\
91.3\%  (SVM for Anger and Neutral),\\
88.5\%  (SVM for Happiness and Anger),\\
81.2\%  (SVM for three emotions)}\\ 
\midrule
\cite{quiroz2018emotion} & \tabincell{ll}{Happiness,Sadness,\\Neutrality }&  \tabincell{llll}{Accelerometer data recorded \\by smart watch:\\50 people divided into \\three task groups experienced \\happiness and sadness;\\Feature vectors extracted \\from sliding windows} & \tabincell{ll}{Random Forest,\\Logistic Regression}& \tabincell{ll}{Accuracies mainly range 60\% to 80\% \\ for three groups}\\ 
\midrule
\cite{chiu2018emotion} & \tabincell{ll}{Joy,Anger,Sadness,\\Relaxation,Neutrality }&  \tabincell{llll}{Estimated joint positions from video: \\11 male walkers experienced \\each emotion respectively;\\Euclidean features,angular features\\ of joints,and speed} & \tabincell{ll}{SVM,\\Decision
Tree,\\Naive Bayes,\\ Random Forest,\\ Logistic Regression,\\Multilayer Perceptron}& \tabincell{ll}{Highest accuracy:\\ 62.1\% using SVM}\\ 
\midrule
\cite{ahmed2018score} & \tabincell{ll}{Happiness,Sadness,\\Anger,Fear,Neutrality }&  \tabincell{llll}{Kinect skeletons:\\7 walkers experienced \\each emotion respectively;\\Geometric and kinematic \\features using LMA framework} & \tabincell{ll}{KNN,SVM,LDA,\\Decision Tree,\\Genetic algorithm,\\Score-level Fusion,\\Rank-level Fusion }& \tabincell{ll}{Highest accuracy:\\80\% using score-level fusion \\of all four models}\\ 
\bottomrule

\end{tabular}}}
 \begin{tablenotes}
        \footnotesize
        \item[1]KNN: K Nearest Neighbor. SVM: Support Vector Machine. LDA: Linear Discriminant Analysis. LMA: Laban Movement Analysis
      \end{tablenotes}

\label{acc_table}
  
\end{table*}

\section{Future Directions}
In this paper, we have gained insight into how emotions may influence on gaits and learned different techniques for emotion recognition. The development of this field has a lot of potentials, especially with the advancement of intelligent computation. Based on the current research, we highlight and discuss the future research directions as below. 
 
\par
{\it Non-contact Capturing.} Different gait-based analyses depend on different devices that capture and measure relevant gait parameters. There are lots of capturing systems and they might be divided into three subsets: wearable sensors attached onto human body, floor sensors and sensors for image or video processing\cite{muro2014gait}. It is no doubt about the validity of these types of sensors used in gait detection and measurement as numerous investigations have already demonstrated. Although wearable sensors are very precise to record the trajectories of motion because of their relatively large number of markers and the high sampling rate, they may lead to uncomfortable feeling and unnatural motion as they have to be directly attached to subject's body. Secondly, force platform requires to be paved on the ground which means it is not advisable to upgrade to large scale due to the expense. On the other hand, as the number of the surveillance cameras increase in various public places such as streets, supermarkets, stations, airports and offices, the advantage of using non-contact gait measuring based on image or video becomes so tremendous. It can achieve big data for training comprehensive models to identify the walkers' emotion in order to avoid the overfitting problem \cite{lecun2015deep}. 
In addition, it is a more natural way to record people's gaits because there is no disturbance to their emotional representation compared to putting markers on their body. Another new technology, depth sensor \cite{zhang2012microsoft} which can be integrated into RGB sensor, has become popular. It is a remarkable way to measure the distance of things with $x,y,z$ coordinates data.  
The studies for identifying emotion on the basis of gait through RGB-D information are still rare, which presumably will become one of breakthrough in the future. \par
{\it Intelligent Classification.} 
From Table \ref{acc_table}, we find that there is no study in the field of gait-based emotion recognition that takes advantage of deep learning techniques, which probably would become one of the future breakthroughs. 
Deep learning algorithms help to achieve great success in artificial intelligence. 
In particular, the convolutional neural networks  (CNN) have gained massive achievements for image-based task and the recurrent neural networks  (RNN) is able to deal with sequence-based problems \cite{wang2018rgb}. As for gait recognition, deep CNNs can help a lot in classification as well as dealing with cross-view variance  \cite{wu2017comprehensive} whereas RNN can tackle with the temporal pattern of the people's gait. Furthermore, the combination of different deep neural networks may yield better outcome since only one single network could not ensure comprehensive solution. For instance, The network  C3D+ConvLSTM is hybrid type that fuses CNN and LSTM together. It can lead to good performance for motion recognition \cite{zhu2017multimodal}. As the gait based emotion recognition is intrinsically a classification task, it should take full advantage of the fusion of various deep neural networks. Furthermore, if the process is supposed to be achieved massively based on RGB or RGB-D information like surveillance in public, there will be some problems to be handled. For example, many subjects will be captured simultaneously in the view which leads to chaos of targeting. The key solution is to do parallel target segmentation \cite{zacharatos2014automatic}. At the same time, the segmentation should be intelligent enough to figure out if the target remains the same identity from the beginning, which still depends on intelligent computation. Apparently, how to make full use of computational intelligence for boosting the classification accuracy is becoming momentous.\par
{\it Large-scale Data Sets.} Along with the development of computation intelligence, big data is desperately required for deep network training in order to facilitate the generalization and robustness of the models. 
To establish a large and comprehensive data set for gait based emotion recognition, several elements should be taken into account. 
Firstly, the number of participants should be large enough. The second point is the variety of participants. 
They should include different genders, wide range of age, various height and weight, with or without carrying things etc. 
The next point would be the emotion elicitation by volunteers. Participants producing approximate emotion expression is quite important because it leads to clear and fine labels for model training. Thus, finding an appropriate way to elicit volunteers' emotion naturally is a significant aspect.  
Getting volunteers to walk in various spatial contexts is another requirement, since walking on different contexts such as on sand or lawn may influence the gait. 
Last but not least, it is advisable to collect data through different views of cameras and in different formats for enriching the information and training more generalized models. The HumanID Gait Challenge data set may act as great template to learn for establishing the emotion-gait related data set. 
It consists of 1,870 sequences from 122 subjects spanning five covariates including change in viewing angle, change in shoe type, change in walking surface, carrying or not carrying a briefcase, and elapsed time between sequences being compared \cite{sarkar2005humanid}. 
These requirements discussed above are not easy to accomplish but are necessary for building a large scale and comprehensive data set.\par
{\it Online Emotion Prediction.} If the requirements above are fulfilled, as an ultimate goal, online emotion prediction based on human gait is possible. However, online prediction requires data analysis to be run in a continuous way. Different from object detection which can be achieved almost real-time \cite{ren2015faster}, online emotion prediction based on gait has to consider the historic steps since the gait cycle has a duration. Therefore, there is a high demand on computation capability. It is usually time-consuming if the neural network is deep, so real-time recognition could be hard to be ensured. 
Thus, to develop more efficient methods for online emotion recognition with less computing resources would be a challenging and important task. 

\section{Conclusions}
This survey paper has provided a comprehensive study on the kinematics in emotional gaits and reviewed emerging studies on gait-based emotion recognition. It has discussed technical details in gait analysis for emotion recognition, including data collection, data preprocessing, feature selection, dimension reduction, and classification. We want to conclude that gait is a practical modality for emotion recognition, apart from its existing functions of motion abnormality detection, neural system disorder diagnosis, and walker identification for security reasons. Gait analysis is able to fill the gaps in automated emotion recognition when neither speech nor facial expression is feasible in long-distance observation. The technique of gait analysis is of great practical value even though it still has a lot of research challenges. Fortunately, there are more and more non-invasive, cost-effective, and unobtrusive sensors available nowadays, like surveillance cameras installed in public, which make the collection of big data possible. Coupling with the development of intelligent computation, gait-based emotion recognition has great potential to be fully automated and improved to a higher level to support a broader range of applications.

\ifCLASSOPTIONcaptionsoff
  \newpage
\fi



%

\bibliographystyle{IEEEtran}
\bibliography{main}

\end{document}